\documentclass[aps,prb,preprint,unsortedaddress,showpacs]{revtex4}
\usepackage{amssymb}
\usepackage{graphicx}
\usepackage{color}

\begin{document}

\title{ Electronic, vibrational and thermodynamic properties of  $\beta$-HgS (metacinnabar), HgSe and HgTe}

\author{M. Cardona}
\author{R. K. Kremer}
\email[Corresponding author:~E-mail~]{r.kremer@fkf.mpg.de}
\author{R. Lauck}
\author{G. Siegle}
\affiliation{Max-Planck-Institut f{\"u}r Festk{\"o}rperforschung,
Heisenbergstrasse 1, D-70569 Stuttgart, Germany}

\author{A. Mu\~{n}oz}
\affiliation{MALTA Consolider Team, Departamento de F\'{\i}sica Fundamental II, and Instituto de Materiales y Nanotecnolog\'{\i}a, Universidad de La Laguna, La Laguna 38205, Tenerife, Spain}

\author{A.H. Romero}
\affiliation{CINVESTAV, Departamento de Materiales, Unidad
Quer$\acute{e}$taro, Quer$\acute{e}$taro, 76230, Mexico}

\date{\today}

\begin{abstract}
We report \textit{ab initio} calculations of the electronic band structure and the phonon dispersion relations of the zincblende-type mercury chalcogenides ($\beta$-HgS, HgSe, and HgTe). The latter have been used to evaluate the temperature dependence of the specific heat which has been compared with experimental data. The electronic band structure of these materials has been confirmed to have an inverted direct gap of the $\alpha$-tin type, which makes HgSe and HgTe semimetallic. For $\beta$-HgS, however, our calculations predict a negative spin-orbit splitting which restores semiconducting properties to the material in spite of the inverted gap. We have calculated the spin-orbit induced linear terms in \textbf{k} which appear at the $\Gamma_8$ valence bands. We have also investigated the pressure dependence of the crystal structure and the phonons.
\end{abstract}

 \maketitle


\email{M.Cardona@fkf.mpg.de}

\section{Introduction}\label{SecIntro}

The present work is part of an on-going effort to investigate, both theoretically and experimentally, the temperature dependence of the specific heat $C_v \approx C_p$  of semiconductors and semimetals in the low temperature region in which strong deviations from the Debye $T^3$ law take place.\cite{Heatpress} These deviations lead to a peak in $C_p/T^3$ vs. $T$  at a temperature $T \approx$ $\Theta_{\rm}$/20, where $\Theta_{\rm}$  is the Debye temperature. Previous investigations have focused on diamond-like semiconductors  and their zincblende and wurtzite analogs.~\cite{Sanati2004,Cardona2005,Gibin2005,Kremer2005,Serrano2006}
More recently the lead chalcogenides, with rock salt structure, have been investigated.~\cite{Romero2008} They, as well as the structurally related semimetals Bi and Sb, exhibit a remarkable dependence of their vibrational properties, and correspondingly $C_v$, on spin-orbit (\textit{s-o}) interaction.~\cite{Diaz2007PRL,Serrano2008} $C_v$ calculations, and the corresponding $C_p$ measurements, have also been reported for the related semiconductor (rhombohedral structure) GeTe,  a material which  grows with a large hole concentration that enables the investigation of the effect of such doping on the vibrational properties and on $C_p$ (or $C_v$).~\cite{Shaltaf2009} This work usually included measurements on samples prepared with different isotopic compositions, so as to investigate their effect on $C_p$ , accompanied by similar calculations of $C_v$.

Because of the calculated dependence of the aforementioned properties on \textit{s-o} interaction for materials composed of heavy atoms (e.g. Bi, Pb) and structures related to rock salt, we decided to investigate similar materials that crystallize in the zincblende structure. The mercury chalcogenides appeared to be ideal for this purpose, although the toxic nature of mercury dissuaded us from investigating isotope effects. Moreover, the seven stable isotopes of Hg have low isotopic abundances and correspondingly prohibitive prices. These materials, however, have a number of hitherto not fully investigated electronic properties, related to the \textit{s-o} interaction and to the gap inversion, which makes of them zero-gap semiconductors.~\cite{Groves1963} We have investigated some of these properties in the process of calculating \textit{ab initio} the phonon dispersion relations and $C_v$.

HgS is usually found in nature as cinnabar, crystallized in the Se-like trigonal structure ($\alpha$-HgS), a structure in which it usually also appears when grown in the laboratory. A small addition  of Fe helps to crystallize HgS in the zincblende structure ($\beta$-HgS), both in nature (($\sim$1\%) Fe), metacinnabar\cite{Zallen1970}) and in the laboratory.~\cite{Szuszkiewicz1995} We use for the present work natural crystals of $\beta$-HgS originating from Mount Diablo Mine, Clayton, CA (purchased from John  Betts, New York, NY 10025).
{\color{black} A chemical microanalysis of two such crystals taken from different parts of the sample gave an Fe content of $\sim$ 1.1 mass-\% Fe.
Additionally, we measured the specific heat of a
commercially available  polycrystalline fine powder  of 99.999\% stated purity (Alfa Aesar GmbH \& Co KG, Karlsruhe).}
HgSe and HgTe can also be found in nature (HgSe: Tiemannite, HgTe: Coloradoite) but they are rare. We used for the work reported here synthetic crystals grown by the Bridgman technique. Cinnabar is transparent in the red and has an optical gap of $\sim$2.275 eV.~\cite{Masse1978} We shall not discuss it here any further since we confine ourselves to the chalcogenides with zincblende structure. It will be discussed in a later publication. The zincblende variety of HgS ($\beta$-HgS, metacinnabar), which will be fully discussed here, has a black color and a zero or near zero gap.
All crystals used here were found to be $n$-type by thermoelectric power measurements.

The structure of this article is as follows:
In Sec. \ref{SecTheory} we discuss the procedures employed for the \textit{ab initio} calculations of the electronic band structure, the phonon dispersion relations and the heat capacity.
Section \ref{SecBandStruc} presents some important results obtained in the process of performing the band structure calculations. We display the full LDA electronic band structure of $\beta$-HgS, which has not been reported in the literature so far. We have found, however, three articles discussing the band structures of HgSe and HgTe which make it unnecessary to reproduce here our full calculated band structures for these materials. The oldest, by Bloom and Bergstresser\cite{Bloom1970}  uses an empirical pseudopotential method which includes \textit{s-o} interaction. The second one, by Overhof~\cite{Overhof1971}, uses the KKR (Korringa-Kohn-Rostocker) method. In spite of not being self-consistent, these calculations already reveal the inverted, $\alpha$-tin like band structure of these materials, as it does the self-consistent LMTO calculation for HgTe performed by Cade and Lee.~\cite{Cade1985} In the process of analyzing our band structure calculations, we realized that there is a considerable contribution of the core 5\textit{d} levels of Hg to the \textit{s-o} splitting at the $\Gamma$ point of the valence bands. Like in the case of the copper monohalides, this contribution has a sign opposite to that of the splitting of the \textit{p} valence electrons of the anion.~\cite{Cardona1963,Shindo1965}  For HgS, this 5\textit{d} contribution overcompensates that of the 3\textit{p} states of sulfur, thus reversing the corresponding valence states and changing the topology of the bands. This situation is similar to that found in CuCl, ZnO and CuGaS$_2$.~\cite{Tell1975}  The calculated \textit{s-o} splittings for the three Hg compounds are given in Table \ref{Table1}. This table also contains calculated values of the negative $\Gamma_8$ - $\Gamma_1$  gap and the lattice parameter
$a_0$  obtained through energy minimization. Experimental data for these parameters are also given.

Another interesting aspect of the band structures of the HgS, HgSe and HgTe is the appearance of linear terms in the wavevector ${\bf k}$ at the $\Gamma_8$ point of the valence bands. Such terms, of relativistic origin, were predicted by G. Dresselhaus   for zincblende-type materials on symmetry grounds.~\cite{Dresselhaus1955} They arise from \textit{s-o} coupling plus the lack of inversion symmetry and involve \textbf{k}$ \cdot $\textbf{p} interactions between the valence bands and the core \textit{d} electrons.~\cite{Cardona1988}  Very few \textit{ab initio} calculations of these terms, which are important in present day spintronics, have been performed and the Hg chalcogenides should be ideal materials for such calculations. We thus present enlarged versions of the valence band structures around $\Gamma$ and use them to evaluate the coefficients of these linear terms. The results are presented in Table \ref{Table2}.

Section \ref{SecLattDyn} is devoted to the \textit{ab initio} lattice dynamical calculations and comparison with inelastic neutron scattering and Raman data. For the purpose of interpreting the specific heat results, we also present the corresponding one-phonon densities of states and their decomposition in Hg and anion contributions. The possible contributions to the linewidths of the Raman phonons, anharmonicity, electron phonon interaction and isotopic disorder is also discussed.
For completeness, we also present the density of two-phonon states (sum and difference) corresponding to a total zero ${\bf k}$-vector.
These results are useful for the interpretation of two-phonon Raman scattering data; they are compared with spectra available in the literature for HgSe.

Because of the lack of theoretical information concerning the phase transitions of the Hg chalcogenides under pressure, we have also performed calculations of the enthalpy of HgS and HgTe versus pressure for the zincblende, rocksalt and cinnabar phases. From the calculated dependence on pressure \textit{p}, the values of \textit{p} at which transitions occur are obtained and compared with experimental results. As a by-product we report calculated values of the Gr\"uneisen parameters of the LO and TO phonons of HgTe at the $\Gamma$-point of the Brillouin zone. This work is presented in Sect. \ref{SecPressEffects}.

Section \ref{SecHeatCap} discusses the specific heats and their temperature dependence around the maximum of $C_v/T^3$. The \textit{ab initio} calculations, which have been performed with and without \textit{s-o} interaction and using either the LDA or the GGA approximation, describe rather well the experimental results. The inclusion of \textit{s-o} interaction lowers the maximum of $C_v/T^3$ by about 8\% in the three compounds with zincblende structure. Since this effect has a sign opposite to that found for the rock salt structure semiconductors (e.g. PbS), we tried to ascertain whether the sign of the effect depends on crystal structure. For this purpose, we calculated the lattice dynamics and $C_v$ of HgS in the pressure region where the rock salt phase is the stable one. We found a very small effect of the \textit{s-o} interaction at the maximum of $C_v/T^3$, corresponding to a decrease by less than 1\%. Section \ref{SecSummary} summarizes the conclusions.

\section{Theoretical Details}\label{SecTheory}

In this paper we study the vibrational and thermal properties of HgX (X=S, Se and Te) by three different approaches: an experimental one and two different calculations based on different implementation of density functional theory.~\cite{dft} For the theoretical part, we have used two different computer codes: ABINIT and VASP.~\cite{abinit,vasp} In order to describe the exchange and correlation effects we have used two well known approximations, LDA (local density approximation)\cite{lda,CA}  and PBE-GGA (Perdew, Burke, Ernzerhof generalized gradient approximation).~\cite{pbe}

In the ABINIT calculations, the wavefunction is expanded in plane waves and a norm-conserving pseudopotential approximation to the free ion potential is employed, with the Teter-Pad\'{e} parametrization for the exchange and correlation.~\cite{hgh,tp} In particular, we did use the Hartwigsen-Goedecker-Hutter HGH pseudopotentials, where scalar-relativistic and SO terms are combined, such that they fully represent nonlocal pseudopotentials of separable form.~\cite{hgh,pseudo1,pseudo2} A 60 Ha  kinetic energy cutoff was applied and a Monkhorst-Pack grid of 6$\times$6$\times$6 was used in order to describe any electronic quantity within the Brillouin zone. Electronic and vibrational properties were checked with respect to these parameters to ensure convergence. The lattice dynamics was calculated by using an implementation of density functional perturbation theory.~\cite{dfpt} Vibrational frequencies and the vibrational density of states were calculated after the diagonalization of the interatomic force constants (IFC). These were used to obtain the thermal properties  within the harmonic approximation. An original grid of 12$\times$12$\times$12 within the Brillouin zone, with a total of 72 \textbf{q} points, was used to calculate the corresponding set of IFCs. Afterwards, a finer grid was employed to determine the thermal properties, obtained by performing a Fourier fit to the calculated IFCs (for example in order to calculate  $C_v$, a set of 32768 \textbf{q} points was used, which was reduced  by symmetries to 897 points).

 The Vienna \textit{ab initio} simulation package (VASP) is  a density functional theory implementation with pseudopotentials used to obtain the electron wave functions [see Ref.~\onlinecite{vasp} and references therein]. The exchange-correlation energy was taken in the local density approximation (LDA) with the Ceperley Alder prescription. Calculations with the generalized gradient approximation (PBE-GGA) were also performed.~\cite{CA,pbe} The projector-augmented wave (PAW) scheme was adopted and the semicore 5$d$ and 6$s$ electrons of Hg were included explicitly in the calculations.~\cite{paw} The set of plane waves used extended up to a kinetic energy cutoff of 370 eV for HgS and 310 eV for HgTe and HgSe. This large cutoff was required in order to achieve highly converged results within the PAW scheme. We used a dense Monkhorst-Pack grid  for BZ integrations in order to ensure highly converged results in the three compounds (to about 1-2 meV per formula unit). We also used an accurate algorithm during the calculations which allowed us to obtain highly  converged forces for  the calculation of the dynamical matrix. At each selected volume, the crystal structure was fully relaxed to its equilibrium configuration through the calculation of the forces on atoms and the stress tensor.~\cite{forces} In the relaxed equilibrium configuration, the forces are less than 0.002 eV/\AA\ and the deviation of the stress tensor from a diagonal hydrostatic form is less than 0.1 GPa.

Highly converged results on forces are required for the calculation of the dynamical matrix using the direct force constant approach (or supercell method).~\cite{supercell} The construction of the dynamical matrix at the $\Gamma$-point of the BZ is particularly simple and involves separate calculations of the forces in which a fixed displacement from the equilibrium configuration of the atoms within the primitive unit cell is considered. Symmetry helps to decrease the number of such independent distortions, thus reducing the computational effort in the study of the analyzed structures considered in this work. Diagonalization of the dynamical matrix provides both the frequencies of the normal modes and their polarization vectors which allow us to identify the irreducible representation and the atomic character of the phonons modes at the  relevant \textbf{k} point. Supercell calculations were carried out in order to obtain phonon dispersion curves. Starting with a 2$\times$2$\times$2 supercell the direct method delivers correct harmonic phonon frequencies at the  special wave vectors of the Brillouin zone which are compatible within the selected supercell size. If the interaction range would cease at the selected supercell size, the corresponding phonon frequencies would provide  a good basis for the interpolation to all wave vectors. The LO/TO splitting cannot be automatically included when using the direct method, but  the dynamical matrix can be supplemented with a non-analytical term which depends on the Born effective charge tensors and the electronic dielectric constant.~\cite{Solo}
The thermodynamic properties of the crystals are, to a large extent, determined by phonon dispersion and  can be satisfactorily described in most of the cases within the quasiharmonic approximation, as discussed in Ref.~\onlinecite{supercell}.

\section{Electronic Band Structure}\label{SecBandStruc}

Figure 1 displays the band structure of $\beta$-HgS as calculated with the ABINIT code using the LDA approximation (see Sect. \ref{SecTheory}) with inclusion of \textit{s-o} interaction. The band structure shown in this figure was calculated with the energy optimized lattice parameter $a_0$ = 5.80 \AA . The \textit{s-o} splitting of the $\Gamma_{15}$ valence bands (-0.18 eV) cannot be resolved in this figure because of the wide energy scale it covers. It will be discussed later in connection with Fig. 2. The wide energy scale, however, allows us to observe the position of the lowest, sulfur 3\textit{s}-like $\Gamma_1$ valence band, which is not affected by \textit{s-o} interaction, at about -12 eV. Above this band, the 5\textit{d} bands of Hg appear. Without \textit{s-o} interaction they split into an upper quadruplet ($\sim$-6eV, $\Gamma_{12}$) and a lower sextuplet ($\Gamma_{15}$). \textit{s-o} interaction does not split the $\Gamma_{12}$ states but it splits the $\Gamma_{15}$ sextuplet into a doublet $\Gamma_7$ and a quadruplet $\Gamma_8$ , the former being above the latter. The splitting shown in Fig. 1 is 2.1 eV , the same as the corresponding atomic splitting of Hg.~\cite{Hermann1963}

\begin{figure}[ht]
\includegraphics[width=8cm ]{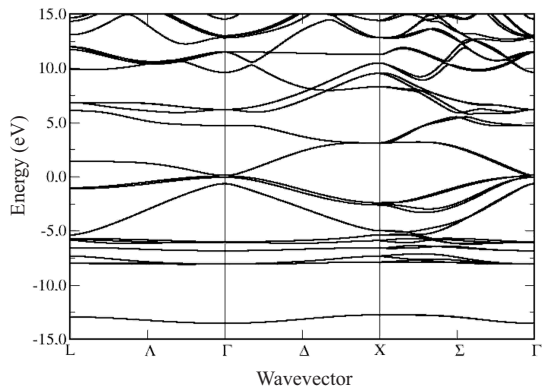}
\caption{Electronic band structure of $\beta$-HgS calculated with the ABINIT code in the LDA approximation (see Sect.\ref{SecTheory}) and with \textit{s-o} interaction using the lattice parameter $a_0$ = 5.80 \AA\  obtained by total energy minimization.} \label{Fig1}
\end{figure}

\begin{figure}[ht]
\includegraphics[width=8cm ]{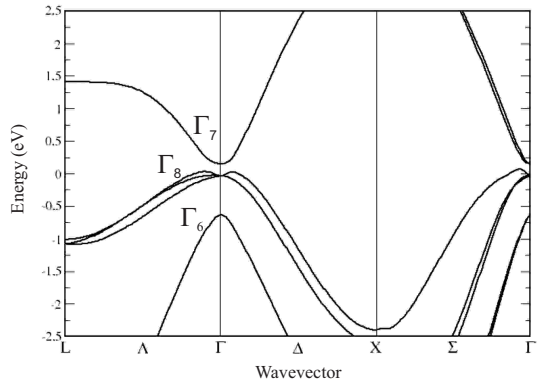}
\caption{Electronic band structure of $\beta$-HgS calculated with the ABINIT code in the LDA approximation (see Sect. \ref{SecTheory}) using the lattice parameter $a_0$ = 5.80 \AA\  obtained by total energy minimization as in Fig. \ref{Fig1}
but with the energy scale enlarged so as to display the details of the top valence bands and the lowest conduction band around $\Gamma$. The $\Gamma_8$ - $\Gamma_7$ separation corresponds to the negative \textit{s-o} splitting $\Delta_0$. $\Gamma_6$ - $\Gamma_8$ is the $E_0$ gap.} \label{Fig2}
\end{figure}

At  -0.58 eV  we find an \textit{s}-like $\Gamma_1$ state which is composed mainly of 6\textit{s} states of Hg. This state corresponds to the lowest conduction band for most tetrahedral semiconductors (exception: $\alpha$- tin~\cite{Groves1963} and, as discussed below, HgSe and HgTe) above the $\Gamma_{15}$ valence bands.
In order to resolve the details of the band structure around the $\Gamma$ point we replot in Fig. \ref{Fig2} the data from  Fig. \ref{Fig1} with an expanded energy scale. This figure shows the inverted \textit{s-o} interaction at $\Gamma$: the $\Gamma_7$ doublet is above the  $\Gamma_8$ quadruplet. This fact arises from the contribution of the negative \textit{s-o} splitting of the 5\textit{d} $\Gamma_{15}$ states of Hg (Fig. 1) which overcompensates the positive splitting of the S 3\textit{p} contribution. This effect is similar to that observed for CuCl (Refs. \onlinecite{Cardona1963,Shindo1965})
but for Hg an additional complication arises, related to the $\alpha$-tin-like structure. Counting bands we see that the $\Gamma_8$ bands should be occupied whereas the \textit{s-o} split $\Gamma_7$ should be empty. Hence, according to Figure 2, undoped (intrinsic) $\beta$-HgS should have an indirect energy gap of $\sim$0.15 eV and a direct one of 0.18 eV . This is in reasonable agreement with the gap observed by Zallen and Slade  (0.25 eV), although these authors did not realize  that the \textit{s-o} splitting of $\beta$-HgS is negative.~\cite{Zallen1970}

\begin{table*}[ht]
\caption{\label{Table1} Values at \textbf{k}=0 of the spin- orbit splitting $\Delta_0$,
the negative \textit{p}-\textit{s} gap $E_0$ (in eV) and the lattice constant $a_0$ (in \AA ) as calculated with various codes and experimentally determined. Notice the negative value of  $\Delta_0$ calculated by our three methods for HgS. It is due to an admixture of \textit{p}-wavefunctions of the cation with 5\textit{d} of Hg (the latter have a negative contribution similar to results found for CuCl, ZnO, CuGaS$_2$). The differences of $\Delta_0$ for the different computational methods are due to minor differences in the compensating \textit{p} and \textit{d} contributions.
The negative $\Delta_0$ of $\beta$-HgS has already been mentioned recently.~\cite{Delin2002,Carrier2004}}
\begin{ruledtabular}
\begin{tabular}{ l  l  c  c  c }
& & HgS & HgSe & HgTe \\ \hline
 & VASP-LDA &	-0.111 &	+0.23&	+0.783\\
& ABINIT-LDA & 	-0.18&	+0.34	& +0.81 \\
$\Delta_0$ (eV) & VASP-GGA &	-0.091 &	+0.25 &	+0.751\\
& Overhof	& - &	+0.48 &	+1.13\\
& experimental	& -0.25 (Ref. \onlinecite{Zallen1970})	& +0.38 (Ref. \onlinecite{Landolt}) &	~0.9 (Ref. \onlinecite{Landolt})\\ \hline
& VASP-LDA &	-0.573 &	-1.18	& -1.025\\
& ABINIT-LDA	&  -0.58	&  -1.26 &	-1.15\\
$E_0$ (eV)& VASP-GGA	& -0.483	& -1.07	& -1.113\\
& Overhof &	- &	-0.4	& -0.3\\
& experimental	& -0.15 (Ref. \onlinecite{Zallen1970}) &	$\sim$-0.4 (Ref. \onlinecite{Landolt}) &  	-0.3 (Ref. \onlinecite{Landolt})\\ \hline
& VASP-LDA &	5.825	& 6.065 &	6.433\\
&ABINIT-LDA	& 5.80	& 6.054	& 6.442\\
$a_0$ (\AA ) &VASP-GGA	& 5.999	& 6.255	& 6.633\\
&ABINIT-GGA	& 5.982	& 6.250	& 6.656\\
& Overhof	& - &	6.075 &	6.45\\
&experimental	& 5.841($\sim$295 K) (Ref. \onlinecite{Rittner1943})&	6.085 (293K) (Ref. \onlinecite{Landolt}) &6.453   (Ref. \onlinecite{Landolt})\\
\end{tabular}
\end{ruledtabular}
\end{table*}

In Table \ref{Table1} we have listed the negative  gaps $E_0$ between the valence bands and the $\Gamma_6$ Hg 5\textit{s}-like conduction band calculated and observed for the three Hg chalcogenides. We notice in the table that the calculated positions of the $\Gamma_6$  bands are lower then the experimental ones by 0.43 eV for HgS and 0.85 eV for HgSe and HgTe. It is natural to attribute this fact to the so called "gap problem" inherent to the LDA calculations.~\cite{Godby1986} 
Recent \textit{GW} show that the $\Gamma_1$ state of the HgX materials should be raised by about 0.75 eV.~\cite{Rohlfing1998,Fleszar2005} If we add this
energy to the values of $E_0$ displayed in Table \ref{Table1}, the $E_0$ gaps of HgSe and HgTe remain negative ($\sim$ -0.35 eV) whereas that of $\beta$-HgS becomes positive ($\sim$ +0.4 eV). Thus $\beta$-HgS would be a conventional semiconductor except for the negative $\Delta_0$.
As already mentioned, we do not reproduce here the full band structures of HgSe and HgTe because they have already been presented in the literature.~\cite{Bloom1970,Overhof1971,Cade1985}  We have given a few important parameters in Table \ref{Table1}, also for HgSe and HgTe. An interesting fact is their positive \textit{s-o} splitting $\Delta_0$ as opposed to the negative one found for HgS. The situation is similar to that found for the copper halides: CuCl has a negative splitting whereas for CuBr and CuI it is positive.\cite{Cardona1963}  This fact has been attributed by Shindo and Kamimura  to the negative contribution of the copper 3\textit{d} electron to the valence band splitting.\cite{Shindo1965} As already suggested above, a similar explanation holds for the Hg chalcogenides: the negative splitting of the 5\textit{d} electrons of Hg contributes, through hybridization, an amount to the valence splitting sufficient to reverse the 3\textit{p} splitting of S but neither the 4\textit{p} splitting of Se nor the 5\textit{d} of tellurium. According to Table I, agreement between experimental and calculated values of $\Delta_0$ is good. Notice that, because of the positive \textit{s-o} splitting, HgSe and HgTe are zero gap semiconductors, like $\alpha$-tin.

In the process of performing these calculations we also took a look at the terms linear in \textbf{k} which appear in the valence bands  around $\Gamma$ as a result of \textit{s-o} interaction coupled with the lack of inversion symmetry. The existence of these terms was first reported  by Dresselhaus \cite{Dresselhaus1955} and numerical calculations were performed by Cardona, Christensen and Fasol\cite{Cardona1988},  by Surh,  Li and Louie\cite{Surh1991} and more recently by Kim \textit{et al.}\cite{Kim2009}. These terms arise in perturbation theory from \textit{s-o} interaction between the valence bands and the \textit{d} core electrons ($\Gamma_{12}$ states) and back to the valence band via \textbf{k}$\cdot$\textbf{p}.\cite{Cardona1988}  In Ref. \onlinecite{Cardona1988} \textit{ab initio} calculations of these terms were presented for many III-V and II-VI zincblende-type semiconductors but not for the Hg-chalcogenides. These calculations may have been vitiated by "gap problems" but concerning the linear terms in \textbf{k} any such errors should be small because they do not involve the \textit{p}-\textit{s} gap. In Ref. \onlinecite{Cardona1988} a semiempirical expression was given for the coefficient $C_k$ of these linear terms. For the Hg-chalcogenides compounds it can be written as (see Eq.(7.6) of Ref. \onlinecite{Cardona1988}):

\begin{equation}\label{eqCk}
C_k = -350 \frac{D_{d,c}}{E(\Gamma_8^{\nu}) - E_{d,c}(\Gamma_{12})} \rm{(in\,\, meV\,\AA )}
\end{equation}

Where $\Delta_{d,s}$ is the \textit{d} \textit{s-o} splitting of the cation (for Hg 2.2 eV) and the energy denominator represents the gap between the $\Gamma_8$ valence band and the outermost \textit{d}-states of the cation. In Eq.\ref{eqCk} we have neglected the contributions of the anion \textit{d} electrons since they are absent for HgS and should be small for HgSe and HgTe. Equation \ref{eqCk} allows us to estimate $C_k$  = -128 meV\,\AA\ for the Hg-chalcogenides.

We now give some details concerning the definition of $C_k$. This parameter has been chosen such that the splitting for \textbf{k} along [100] is $\pm C_k \, k$ (the convention used for the sign is that of Ref. \onlinecite{Cardona1988}), each one of the two split states being a doublet for this \textbf{k} direction.
For \textbf{k} along [111] two of the four $\Gamma_8$ states are not affected by linear \textbf{k} terms, whereas the other two split like $\pm \sqrt{2} C_k \, k$ . For \textbf{k} along [110] the $\Gamma_8$ states split into 4, two like $\pm C_k\, k(1 + \sqrt{3}/2)^{1/2}$
and the other two like $\pm C_k\, k(1 - \sqrt{3}/2)^{1/2}$.~\cite{Dresselhaus1955,Surh1991}

\begin{figure}[ht]
\includegraphics[width=8cm ]{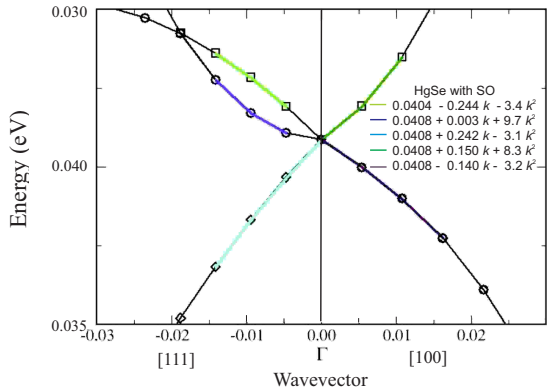}
\caption{(colour online) $\Gamma_8$ valence bands of  HgSe showing the linear terms in \textbf{k} along the [100] and the [111] directions, as calculated with the ABINIT-LDA code. The colored curves represent fits to a quadratic polynomial, the black ones connect calculated points as a guide to the eye. The values of the fit coefficients are given in the inset. The zero of energy lies at the valence band maximum at the Fermi energy.} \label{Fig3}
\end{figure}

We display in Fig. \ref{Fig3} the valence bands of HgSe calculated in a small energy and k range so as to be able to observe the linear terms and thus determine $C_k$. The calculations were performed for \textbf{k} along [100] and [111] using the ABINIT code in the LDA approximation. Values of $C_k$ were obtained through fits with a polynomial containing a linear and a quadratic (effective mass) term. Notice the splitting pattern mentioned above: for \textbf{k} along [100] the $\Gamma_8$ quadruplet splits into two doublets. For \textbf{k} along [111] a doublet does not split and exhibits no linear term (notice the small value of the linear fitting coefficient). The other doublet splits into $\pm \sqrt{2} C_k\,k$. Similar patterns are found for HgS and HgTe. We shall not present here  the corresponding figures. Instead we shall list all of the values of $C_k$ determined for our three materials with the ABINIT and the VASP codes.

We display in Table \ref{Table2} the values of $C_k$  for the three materials as found with the two computational codes from the [100] and the [111] splittings. The sign convention is the same as that of Ref. \onlinecite{Cardona1988} and Ref. \onlinecite{Surh1991}.  These values are rather close to that estimated with
Eq. \ref{eqCk} using for the \textit{p}-\textit{d} gap 6 eV and 2.2 eV for the \textit{s-o} splittings of the 5\textit{d} electrons of Hg  ($C_k$ = -128 meV\,\AA). We have not found experimental values of $C_k$
for these materials.

\begin{table*}
\caption{\label{Table2} Values of the coefficients of the linear terms around the $\Gamma_8$ states (in meV\,\AA\ ) of the valence band structures of $\beta$-HgS, HgSe and HgTe as calculated with the VASP and ABINIT codes
through a quadratic fit of the eigenvalues (see Fig. 3).}
\begin{ruledtabular}
\begin{tabular}{ccccccc}
& \multicolumn{2}{c}{$\beta$-HgS}& \multicolumn{2}{c}{HgSe}  &
 \multicolumn{2}{c}{HgTe} \\
 \cline{2-3}\cline{4-5}\cline{6-7}
         & $C_k$ [100] & $C_k$ [111] &$C_k$ [100]&  $C_k$ [111] &$C_k$ [100]&  $C_k$ [111] \\ \hline
 VASP- GGA   & -147 & -140 & -145 & -146 & -145 & -137\\
   ABINIT-LDA & -144 & -174 & -160 & -170 & -148 & -136 \\
\end{tabular}
\end{ruledtabular}
\end{table*}

\section{Lattice Dynamics}\label{SecLattDyn}

\begin{figure}[ht]
\includegraphics[width=8cm ]{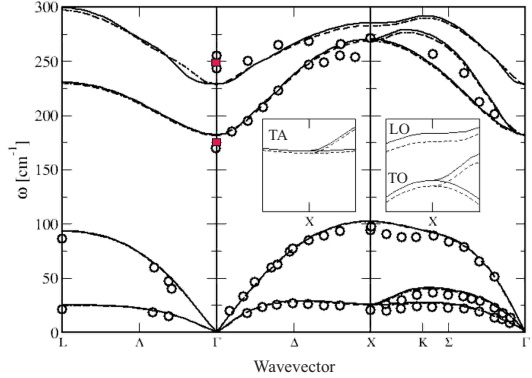}
\caption{(color online) Phonon dispersion relations of $\beta$-HgS calculated with the ABINIT-LDA code with (solid line) and without (dashed line) \textit{s-o} coupling. The circles were obtained with INS~\cite{Szuszkiewicz1998}.  The LO and TO phonons at \textbf{k} $\sim$ 0 (red) squares were obtained by Raman scattering.\cite{Szuszkiewicz1995}
The frequency scales in the insets have been expanded to show the effect of the \textit{s-o} around X.} \label{Fig4}
\end{figure}

A few Inelastic Neutron Scattering (INS) determinations of the phonon dispersion relations of $\beta$-HgS~\cite{Szuszkiewicz1995Acta,Szuszkiewicz1998},  HgSe~\cite{Kepa1980} and HgTe~\cite{Kepa1980}  have been published. They have been compared with semiempirical calculations which use fitted force constants. Here we compare the measured points with \textit{ab initio} calculations (ABINIT or VASP) and examine the effect of \textit{s-o} interaction on these calculations.
We start with $\beta$-HgS and display in Fig. \ref{Fig4} the phonon dispersion relations calculated with \textit{s-o} interaction (solid line) and also without (dashed line). The effect of the \textit{s-o} interaction (see inset for the TA phonons) is rather small, however it is possible to see that for the TA phonons, those mainly responsible for the $C_v$ to be discussed later, the frequencies without \textit{s-o} are about 5\% lower than with \textit{s-o}.

\begin{figure}[ht]
\includegraphics[width=8cm ]{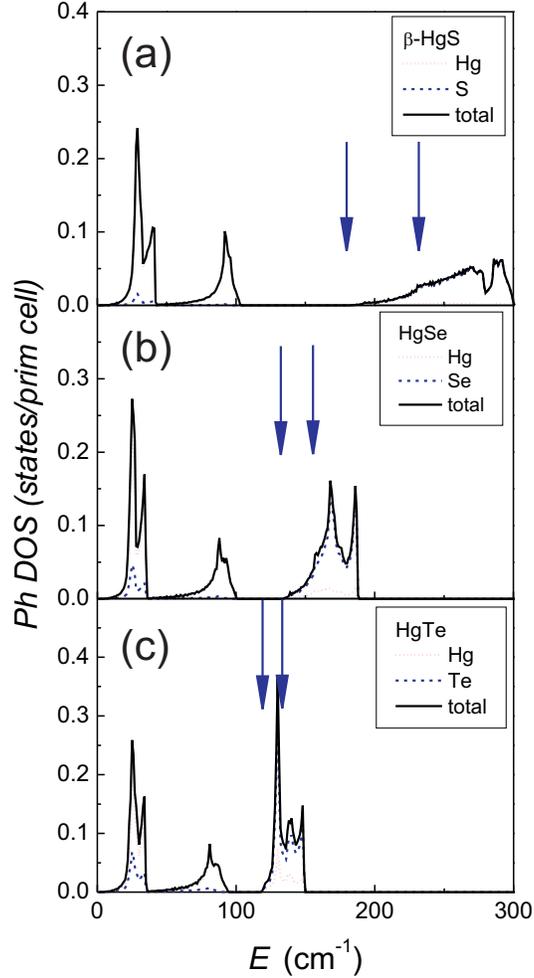}
\caption{(Colour online) Phonon density of states of $\beta$-HgS, HgSe and HgTe, (a), (b), and (c), respectively, as obtained with the ABINIT code used for Fig. \ref{Fig4}.
We have plotted the total DOSs and also their projection on the Hg and the anion atoms. Note that in the case of  $\beta$-HgS the acoustic phonons correspond to nearly pure vibrations of Hg, the optic ones to nearly pure S vibrations.  The (blue) vertical arrows mark the positions of the optical phonons at $\Gamma$.
} \label{Fig5}
\end{figure}

An interesting feature of Fig. \ref{Fig4} is the strong upwards bending of the optical bands away from \textbf{k} = 0, the opposite of what happens for most tetrahedral semiconductors. This effect is due to the large mass difference of Hg and S: at the edge of the zone the "effective" mass which determines the frequency is basically that of sulfur, whereas at \textbf{k} = 0 it is the somewhat smaller  reduced mass. In order to obtain the calculated up-bending one must also invoke a strong force constant connecting the anions. The upward bending is reflected in the corresponding density of phonon states, shown in Fig. \ref{Fig5}, together with those of HgSe and HgTe to be discussed later. A direct consequence is that the TO Raman modes do not overlap the DOS continuum while the LO mode does. The width of the former should therefore not be affected by elastic scattering due to defects, such as isotopic fluctuations, whereas the LO-modes, overlapping the background, should. This situation is opposite to that discussed in Ref. \onlinecite{Goebel1999} for ZnSe, where isotopic disorder is shown to broaden the TO Raman phonons but not the LO phonons.

The up-bending of the dispersion relations   are also important when considering confinement in quantum dots. These effects have been discussed for the case of CdS, a material for which the the TO bulk bands bend up but the LO bands bend down.\cite{Chamberlain1995} No such experiments are available, to the best of our knowledge, for $\beta$-HgS but quantum confinement in dots of this material should produce an increase in frequency as opposed to most other related materials. Up-bending is also found in the dispersion relations of HgSe and HgTe (see below), but not as strong as in the case of  $\beta$-HgS because the difference between anion and cation masses is not as large. A canonical example of up-bending phonon dispersion relations is found in rock salt structure PbS. In this material the LO-phonons are found at 205 cm$^{-1}$ at the $\Gamma$ point of the BZ. A direct consequence of the up-bending is the fact that the corresponding peak in the second order Raman spectrum is not centered at 2$\times$105= 210 cm$^{-1}$ but somewhat higher (228 cm$^{-1}$ at 12K).\cite{Etchegoin2008} In contrast, for GaAs, a typical down-bending case, the two LO-phonon peak is shifted down by as much as 5 cm$^{-1}$ with respect to twice the frequency of the LO phonons at $\Gamma$.~\cite{Olego1981} Effects of the upwards bending on the phonon line widths will be discussed below.

\begin{figure}[ht]
\includegraphics[width=8cm ]{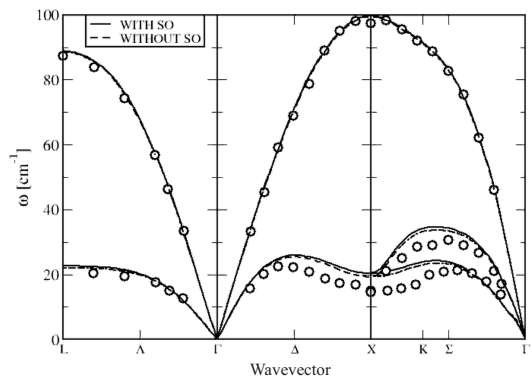}
\caption{\textit{ab initio} calculations of the acoustic dispersion relations of HgSe performed with the ABINIT-LDA code, with (solid line) and without \textit{s-o} (dashed line) interaction. The circles represent INS data taken from Ref. \onlinecite{Lazewski2003}. We also performed calculations with the VASP-GGA code. The LA modes were found to be slightly lower in frequency (82 cm$^{-1}$ at L, 91 cm$^{-1}$ at X) as expected from the larger $a_0$ (see Table \ref{Table1}). The TA modes barely changed.} \label{Fig6}
\end{figure}

\begin{figure}[tbph]
\includegraphics[width=8cm ]{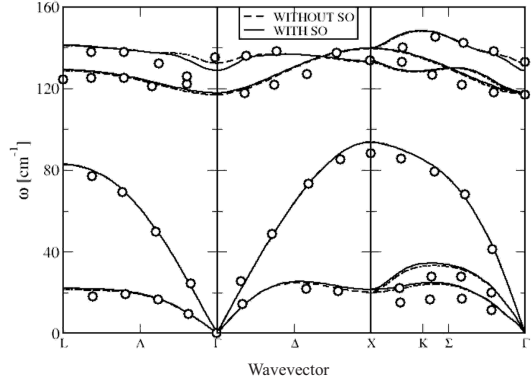}
\caption{Phonon dispersion relations calculated with the ABINIT-LDA code for HgTe. The circles represent data from INS measurements.~\cite{Kepa1980}
Raman spectroscopy gives TO($\Gamma$) = 116 cm$^{-1}$ and  and LO($\Gamma$) = 139 cm$^{-1}$.\cite{Mooradian1973}}\label{Fig7}
\end{figure}

Figures \ref{Fig6} and \ref{Fig7} display the phonon dispersion relations of HgSe and HgTe, respectively, as calculated with the ABINIT-LDA code. In Fig. \ref{Fig6} we only show the acoustic phonon branches, and the corresponding INS data  because \textit{ab initio} calculations, using the VASP-ASA code, have already been reported by {\L}azewski and coworkers.~\cite{Lazewski2003} The main purpose of this figure is to visualize the effect of the \textit{s-o} interaction which slightly raises the frequencies of the TA modes and, as we shall see below, lowers the calculated maximum of $C_v/T^3$.

The optical branches calculated in Ref. \onlinecite{Lazewski2003} for HgSe  also bend upwards when moving away from $\Gamma$, but not  as much as in Fig. \ref{Fig4}. This is expected because for HgSe anion and cation masses differ less than in the case of $\beta$-HgS. Correspondingly, and as shown in Fig. \ref{Fig5}, the optical phonon bands of HgSe are narrower than those of $\beta$-HgS. Moreover, the projected densities of states of these phonons correspond about 90\% to Se and 10\% to Hg.

Figure \ref{Fig7} displays the full acoustic and optic dispersion relations for HgTe together with INS data. In this case the upward bending of the optical bands is rather small, as corresponds to cation and anion masses which differ by less than a factor of two. Because of the weak upward bending, the optical DOS band is even narrower than for HgSe (see Fig. \ref{Fig5}).

For the sake of completeness we include a figure with the two-phonon DOS which applies to second order optical processes in which either two phonons are created (sum processes) or one is created and the other destroyed (difference processes). After multiplication by the appropriate matrix elements (which we have not calculated in this work) the two phonon DOS should describe the corresponding IR and Raman spectra. Even  without multiplication by the matrix elements, structure in the two-phonon DOS should help to interpret the corresponding spectra.

\begin{figure}[tbph]
\includegraphics[width=8cm ]{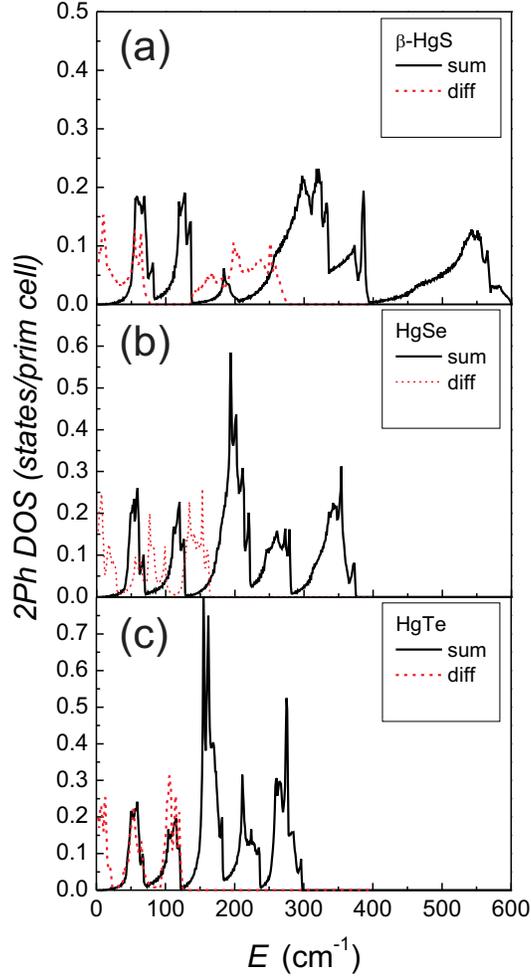}
\caption{(color online) Sum ((black) solid line) and difference ((red) dashed line) of the two-phonon DOS of $\beta$-HgS, HgSe and HgTe. The wavevectors of the two phonons are taken to be equal (but opposite), as required by optical spectroscopies.} \label{Fig8}
\end{figure}

There is only scarce information about the second order Raman and IR spectra of the materials under consideration. We have already mentioned the 2LO structure seen for $\beta$-HgS, which is slightly upshifted with respect to twice the frequency of the LO ($\Gamma$) phonons. Szuszkiewicz \textit{et al.}  have reported Raman spectra for HgSe with structures which can be attributed to two phonon sums.~\cite{Szuszkiewicz1999}   A rather detailed assignment is now possible using the DOS of Fig. \ref{Fig8}. The resolution of their measurements is better than 2 cm$^{-1}$. The FWHM of the observed TO phonon is $\sim$5cm$^{-1}$. After correcting for instrumental resolution~\cite{Kielkopf1973}, the FWHM becomes $\sim$4.2 cm$^{-1}$. We shall discuss below a possible isotope disorder contribution to this width.~\cite{Cardona2005RMP} The temperature at which the spectra were measured (90K) most likely precludes the observation of difference processes. The measurements were performed in back-scattering using a [110] surface of cleavage. Under these conditions, the selection rules prevent the observation of single LO phonons. The single TO mode is seen at 133 cm$^{-1}$. We show in Table \ref{Table3} the frequencies of the observed "two phonon" peaks and their assignment to structures in Fig. \ref{Fig8}. We also have made an attempt, as given in  Table \ref{Table3}, to assign these structures to the dispersion relations calculated for HgSe. The exact
overtones of the phonons at $\Gamma$  [\,2\,TO($\Gamma$)= 264 cm$^{-1}$ and 2\,L0($\Gamma$)=330 cm$^{-1}$\,] are not clearly observed:
2\,TO($\Gamma$) corresponds to a threshold  and not a peak (see Fig. \ref{Fig8}), 2\,LO($\Gamma$) should be slightly shifted upwards because of the up-bending and may overlap with the broad band that we have assigned to 2\,LO(X), as indicated in Table \ref{Table3}.

\begin{table*}
\caption{\label{Table3} Structures observed in the second order Raman spectrum of HgSe (Ref. \onlinecite{Szuszkiewicz1999})  compared with structures seen in the two-phonon DOS calculated on the basis of the band structure partly depicted (only the acoustic bands) in Fig. \ref{Fig7}.}
\begin{ruledtabular}
\begin{tabular}{ccc}
experiment (cm$^{-1}$)  & calculated (cm$^{-1}$)  & assignment \\ \hline
43 &	50, 65 &	2\,TA\\
110 &    110 &	LA + TA\\
120 &    120 &	LA + TA\\
125 &    130 &	LA + TA\\
160 (weak shoulder) & - & -\\
195 &199, 200, 210, 220& O + TA\\
260 & 260, 245, 275, 280  &  2\,TA, LO + TA\\
350  &  230 (exactly 2\,LO($\Gamma$))  &  2\,LO close to $\Gamma$, 2\,LO(X)\\
370 & 370  & 2\,LO (L,$\Sigma$)\\
\end{tabular}
\end{ruledtabular}
\end{table*}

\subsection{
Line widths of $\Gamma$-phonons and isotope disorder contribution to their broadening}

We have found reliable data for these linewidths in the cases of HgSe~\cite {Szuszkiewicz1999} and HgTe~\cite{Mooradian1973}.  The resolution reported in Ref. \onlinecite{Mooradian1973} for HgTe is 1 cm$^{-1}$ (FWHM). Correspondingly, no resolution correction is needed for the measured linewidths which are 5 cm$^{-1}$ for the TO ($\Gamma$) phonon and 4.5 cm$^{-1}$ for LO ($\Gamma$), at 1.6 K.~\cite{Kielkopf1973} In the work of Szuszkiewicz on HgSe the resolution is quoted to have been better than 2 cm$^{-1}$, standard deconvolution~\cite{Kielkopf1973}  of the observed line width of the TO phonons (5 cm$^{-1}$) brings the measured FWHM to about 4.2 cm$^{-1}$. These materials being semimetallic, there should be some contribution of transitions from the occupied to the empty valence bands to the phonon linewidths. An investigation of this contribution has not been performed. The temperature dependence due to anharmonic decay into two phonons~\cite{Debernardi2000}, has not been investigated for these materials, neither experimentally nor theoretically. Nevertheless extant data, even those obtained at the lowest temperatures, are affected by anharmonic effects related to the zero-point vibrations.~\cite{Cardona2005RMP} These zero point effects, however, are typically less than 1 cm$^{-1}$ (FWHM) for most zincblende-type semiconductors.~\cite{Debernardi2000} We thus surmise that the electron phonon coupling across the $\Gamma_8$ valence bands  by either the LO or the TO $\Gamma$-phonons must give a sizeable contribution to their FWHM. Before we assign to electron-phonon coupling  the difference between the observed FWHM ($\sim$5 cm$^{-1}$) and the expected anharmonic effect ($\sim$1 cm$^{-1}$) we must, however, estimate the contribution of the isotopic disorder in materials grown from the natural elements. This can be done with the expression:~\cite{Cardona2005RMP}

\begin{equation}\label{eq41}
\Gamma_i = \frac{\pi \omega_0^2}{6}[\,g_{\rm Hg}\,N_{d{\rm Hg}}(\omega_0)\,|e_{0{\rm Hg}}|^2 + g_{{\rm Ch}}\,N_{d{\rm Ch}}(\omega_0)\,|e_{0{\rm Ch}}|^2]
\end{equation}

where $\Gamma_i$ represents the contribution of isotopic disorder to the FWHM of either the
TO or the LO phonon, $g_{\rm Hg}$  and $g_{\rm Ch}$   the isotopic mass fluctuation parameter of either Hg or the chalcogen, respectively:

\begin{equation}\label{eq42}
g_{\rm Hg,Ch} = \frac{\langle M^2_{\rm Hg,Ch}\rangle - \langle M_{\rm Hg,Ch}\rangle^2}{\langle M_{\rm Hg,Ch}\rangle^2}
\end{equation}

In Eq. \ref{eq41} $N_{d{\rm Hg}}$ and $N_{d{\rm Ch}}$ are the densities of one-phonon states projected on Hg and the chalcogen, respectively, while $e_{0{\rm Hg}}$ and  $e_{0{\rm Ch}}$ are  the corresponding eigenvector amplitudes for the phonons at $\Gamma$.  We list in Table \ref{Table4} the values of $g$ for the natural elements under consideration and for artificial isotopic mixtures conceived so as to maximize $g$ taking into account the price of the isotopes.

\begin{table*}
\caption{\label{Table4} Isotopic fluctuation parameter $g$ ($\times$ 10$^5$) for the elements under consideration, either as found in nature or prepared as equal mixtures of the two isotopes whose masses are given in brackets.}
\begin{ruledtabular}
\begin{tabular}{ccccc}
isoptope composition  & Hg  & S & Se & Te \\ \hline
natural	& 6.2	&  9.2 &	7.9 &	3.8\\
50-50\%	& (198-204)  22	& (32-34)     92 &	(76-80)     67	& (124-130)  56\\
\end{tabular}
\end{ruledtabular}
\end{table*}

\begin{table*}
\caption{\label{Table5} Parameters needed to evaluate the disorder induced broadening of LO ($\Gamma$) phonons with
Eq. \ref{eq41} and the isotope disorder induced broadenings obtained by this procedure.}
\begin{ruledtabular}
\begin{tabular}{ccccccc}
& $N_d{{\rm Hg}}$ & 	$N_{d{\rm Ch}}$ &	$e_{0{\rm Hg}}$ &	$e_{0{\rm Ch}}$	 & $\Gamma_i$ (natural)(cm$^{-1}$) & $\gamma_i$ (50-50\%)(cm$^{-1}$) \\ \hline
$\beta$-HgS	& 0 & 	0.015 &	0.15	& 0.99	& 0.02 &	0.2 \\
HgSe &	0.03 &	0.085	& 0.40 &	0.93	& 0.06 &	0.5\\
HgTe& 	0.02 &	0.05	& 0.53	& 0.85 &	0.02 &	0.2\\
\end{tabular}
\end{ruledtabular}
\end{table*}

We thus conclude that the observed FWHM of TO($\Gamma$) and LO($\Gamma$) phonons in HgSe and HgTe, between 3 and 4 cm$^{-1}$ after subtracting anharmonic effects, must be due to electron-phonon interaction between full and empty valence bands. The effect can be compared to that observed in heavily doped Si and Ge. For Si with 6$\times$10$^{19}$ cm$^{-3}$ holes a contribution of about 5 cm$^{-1}$ to the FWHM is found.~\cite{Chandrassekhar1980}

\section{Pressure Effects on the Structure and Lattice Dynamics}\label{SecPressEffects}

This investigation was triggered by our interest in calculating the specific heats of the Hg chalcogenides crystallized in the rock salt structure (see next section). These materials undergo phase transitions at relatively low pressures.~\cite{Mujica2003} While there is considerable experimental data on these transitions (see Ref. \onlinecite{Mujica2003}), no \textit{ab initio} calculations seem to have been reported. We have performed enthalpy minimizations vs. pressure for  HgS and HgTe considering the three most conspicuous structures of these materials: cinnabar, zincblende and rock salt.  Orthorhombic and bcc structures have also been observed experimentally for HgTe but we have not included them in our calculations.

\begin{figure}[th]
\includegraphics[width=8cm ]{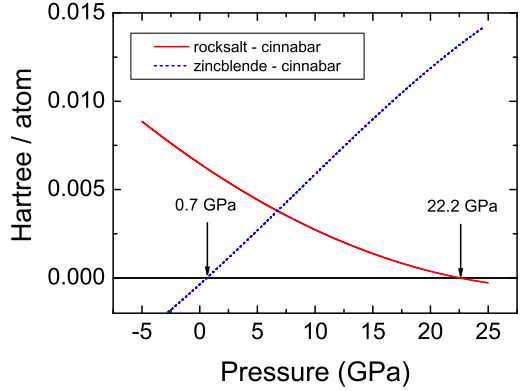}
\caption{(color online)Enthalpy of the three phases of HgS calculated with the ABINIT-LDA code with \textit{s-o} interaction. Cinnabar and rock salt phases are basically degenerate (within any reasonable error estimates) at $p$=0 although formally their enthalpies cross at $\sim$ 0.7 GPa. The cinnabar $\rightarrow$  rock salt transition is predicted to occur at $\sim$22 GPa. Experimental data place it at 20.5 GPa.~\cite{Nelmes1998}} \label{Fig9}
\end{figure}

We show in Fig. \ref{Fig9} the pressure dependence of the enthalpy of the three phases mentioned, as calculated for HgS. The cinnabar phase ($\alpha$-HgS)  is found, within estimated error, to have the same enthalpy as the zincblende phase ($\beta$-HgS), a fact which may explain the ease to obtain both phases in the laboratory, although the growth of the zincblende phase requires the addition of a small amount of Fe.~\cite{Szuszkiewicz1995}

Notice, however, that the zincblende phase is calculated to be the stable one at zero pressure, but with an enthalpy only about $\sim$0.0003 Ha below the cinnabar phase. We do not believe that this minute enthalpy difference is significant. The cinnabar phase is calculated to be stable between 0.7 GPa and 22.2 GPa, at which pressure the NaCl phase should appear. This is in reasonable agreement with reports of a phase transition at 20.5 GPa.~\cite{Nelmes1998}  In the next section we shall present calculations of the specific heat of the rock salt phase of HgS in the stable pressure region, so as to ascertain the effect of \textit{s-o} interaction  on this phase.

\begin{figure}[tbph]
\includegraphics[width=8cm ]{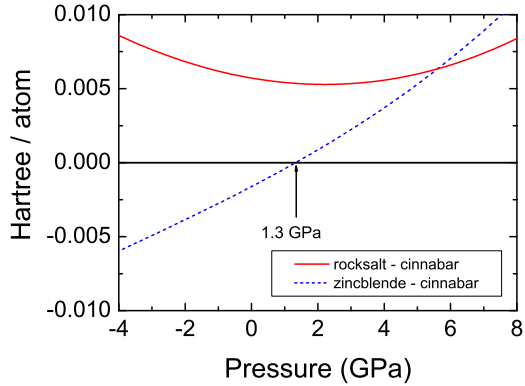}
\caption{Enthalpy of HgTe calculated with the ABINIT-LDA code including \textit{s-o} splitting. The arrow signals the calculated zincblende $\rightarrow$ cinnabar transition which is found experimentally to take place at 1.5 GPa.~\cite{Miguel1995}} \label{Fig10}
\end{figure}

In Figure \ref{Fig10} we present similar calculations for the enthalpy of HgTe vs. pressure.

The calculations were also performed without \textit{s-o} interaction and we found that the presence of this interaction slightly improves agreement with experimental results which are 1.5 GPa for the zincblende $\rightarrow$ cinnabar transition (as compared with 1.3 GPa in Fig. \ref{Fig10}, 1.0 without \textit{s-o} interaction. 8 GPa for the cinnabar $\rightarrow$ rock salt transition has been also observed.~\cite{Miguel1995} For reasons unbeknownst to us it is not reproduced in our calculations.

\begin{figure}[tbph]
\includegraphics[width=8cm ]{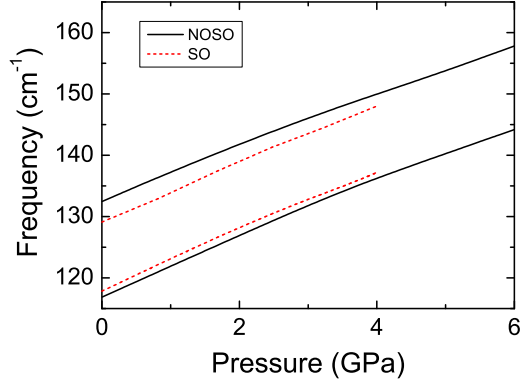}
\caption{Pressure dependence of the frequencies of the LO($\Gamma$) (upper curves) and TO($\Gamma$) phonons (lower curves) of HgTe calculated with the ABINIT code with and without \textit{s-o} interaction as indicated.} \label{Fig11}
\end{figure}

As a complement to this pressure investigation we calculated the pressure dependence of the mode frequencies of the LO($\Gamma$) and TO($\Gamma$) phonons for zincblende HgTe. The results, as obtained with the ABINIT-LDA code with and without \textit{s-o} interaction,  are shown in Fig. \ref{Fig11}

They can be expressed in terms of the usual Gr\"uneisen parameters using as a bulk modulus
$B$= 42.2 GPa:~\cite{Landolt}

\begin{equation}\label{eq51}
\gamma_{\rm LO} = -\frac{d ln\,\omega_{\rm LO}}{d\,ln V} = 1.0;\,\,\,\, \gamma_{\rm TO} = -\frac{d ln\,\omega_{\rm TO}}{d\,ln V} = 1.22
\end{equation}

using Fig.\ref{Fig11} it is also possible to determine a Gr\"uneisen parameter for the transverse effective charge e* as defined with the equation:

\begin{equation}\label{eq52}
\omega_{\rm LO}^2 - \omega_{\rm TO}^2 = \frac{4\pi N e^{*2}}{M \epsilon_{\infty}}
\end{equation}

The calculation of a Gr\"uneisen parameter (i.e. a logarithmic derivative with respect to the volume), like in Eq. \ref{eq51}) for $e^*$ from the data of Fig. \ref{Fig11} requires, beside the bulk modulus, the knowledge of the Gr\"uneisen parameter  $\gamma_{\epsilon_{\infty}}$ of the electronic dielectric constant $\epsilon_{\infty}$.~\cite{Anastassakis1998} For lack of better information we take a value of $\gamma_{\epsilon_{\infty}}$ = +1 , typical of that found for most tetrahedral semiconductors~\cite{Goni1990}  and find for the Gr\"uneisen parameter of $e^*$:  $\gamma^*$= -0.65. This negative sign implies that $e^*$ increases with increasing volume. It is usually interpreted to signal an increase in ionicity with increasing volume, a result that is common to most zincblende-type semiconductors with the exception of SiC (see Table IV in Ref. \onlinecite{Anastassakis1998}).

\section{Specific Heat}\label{SecHeatCap}

The specific heat $C_p$ of the crystalline samples was measured for the three chalcogenides between 2 and 280 K with a PPMS system (Quantum Design, 6325 Lusk Boulevard, San Diego CA (Quantum Design, 6325 Lusk Boulevard, San Diego, CA), as described before.\cite{Cardona2005,Gibin2005,Kremer2005}
{\color{black} The powder samples of $\beta$-HgS were sealed into glass flasks under $\approx$ 900 mbar $^4$He gas to ensure thermal contact at low temperatures. The heat capacities of such samples were measured using a quasi-adiabatic heat pulse method with a home-built Nernst calorimeter similar to that described Schnelle and Gmelin.\cite{Schnelle2002} The heat capacities of the glass sample container, the sapphire platform and a minute amount of Apiezon N vacuum grease used to thermally couple the sample to the platform were measured separately and subtracted.}

\begin{figure}[tbph]
\includegraphics[width=8cm ]{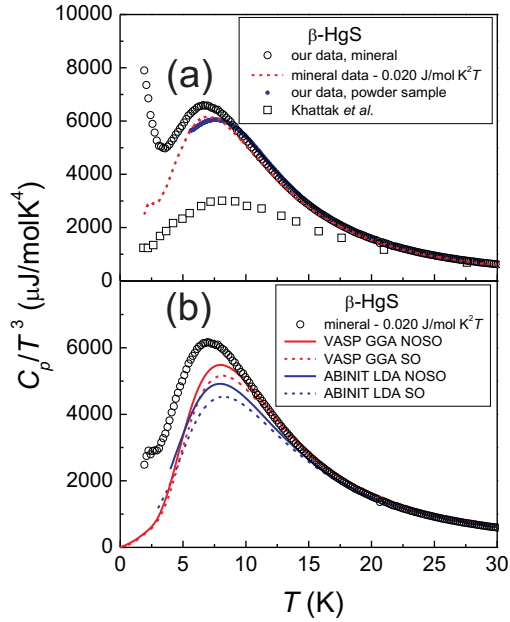}
\caption{(color online) (a) $C_v/T^3$ measured on a mineral and a powder sample of  $\beta$-HgS. Data by Khattak \textit{et al.} are also given for comparison.~\cite{Khattak1981} The solid line represents the mineral data minus a linear contribution as indicated in the inset.
(b) Same mineral data as before compared with the results of four calculations as indicated in the inset.} \label{Fig12}
\end{figure}

{\color{black}
The specific heat of HgSe above 80 K has been measured by Gul'tyaev and Petrov~\cite{Gultyaev1959}.
Low temperature specific heats of HgTe have been determined by Collins \textit{et al.} and subsequently for  $\beta$-HgS, HgSe and HgTe by Khattak and coworkers.~\cite{Collins1980,Khattak1981}
Our data for HgSe and HgTe agree very well with those measured previously.
We find, however, very large differences   at temperatures  around 7.5 K  where the maximum in $C_p/T^3$ occurs for our $\beta$-HgS data compared to those of Khattak \textit{et al.}, our data being  a factor $\sim$2 larger.

Calculations of $C_v$ were performed by integrating the calculated phonon DOS with the standard expression (we have already mentioned that in the region of interest $C_p \approx C_v$). The calculations were performed both, with the ABINIT-LDA and the VASP-GGA codes, with and without \textit{s-o} interaction. The effect of \textit{s-o} interaction was basically the same for the two codes: the \textit{s-o} interaction lowered the maximum of  $C_v/T^3$ by about 10\%. We shall present here the results obtained with only one of the codes (except for $\beta$-HgS) but both, with and without the \textit{s-o} interaction.

We show in Fig. \ref{Fig12}(a) the experimental values of $C_v/T^3$ for two sample of $\beta$-HgS, a natural crystal and a high-purity powder. The natural crystal was found by chemical microanalysis to have an Fe content of 1.1 mass-\%. For comparison we also display measurements by Khatak and coworkers.~\cite{Khattak1981}
The upturn at low temperatures observed for the natural crystal can be ascribed to a linear term in the specific heat (Sommerfeld term) due to metallic behavior. To correct for this term we subtracted a linear contribution $\gamma$=20 mJ/molK$2$) from the experimental data.
Corrections for an FeS content using the data by Kobayashi \textit{et al.}~\cite{Kobayashi1999}
lead to a negligible change of the data and is not shown here.

Fig. \ref{Fig12}(b) displays
calculated VASP and ABINIT values of $C_v/T^3$ for $\beta$-HgS obtained with
and without \textit{s-o} interaction.
The agreement between theoretical and experimental data is reasonably good.

}

\begin{figure}[tbph]
\includegraphics[width=8cm ]{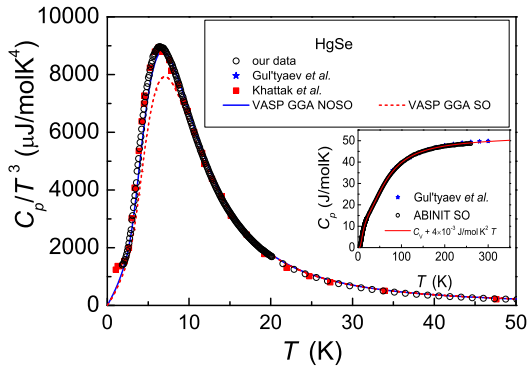}
\caption{(color online) $C_v/T^3$ calculated for HgSe vs $T$ with the VASP code (with (red dashed line) and without (blue solid line) ( \textit{s-o} interaction) and (insert) with the ABINIT code (only with \textit{s-o} interaction). The effect of the latter is similar to that found for the VASP calculations). Also, $C_p/T^3$  measured by us and measurements by Gul'tyaev and Petrov and Khattak \textit{et al.} are presented.~\cite{Gultyaev1959,Khattak1981}
The inset depicts $C_v$$\approx$ $C_p$ obtained by ABINIT with \textit{s-o} interaction included. A contribution linear in temperature, 4$\times$10$^{-3}$ J/mol\,K$^2$ $T$, due to thermal expansion has been added to the calculated data.
} \label{Fig13}
\end{figure}

Figure \ref{Fig13} shows $C_p/T^3$ measured by us and the values calculated from data by Gul'tyaev and Petrov~\cite{Gultyaev1959}, compared with  ABINIT- LDA calculations  which include \textit{s-o} coupling and also with VASP-GGA calculations with and without \textit{s-o} interaction. The latter corroborate the general trend that \textit{s-o} interaction decreases $C_v$ by about 10\%, a fact which corresponds to an increase in the TA phonon frequencies. The VASP calculations agree better with the experiment than the ABINIT ones but the excellent agreement
with experimental data found for the VASP calculations without \textit{s-o} interaction should, of course, be considered fortuitous. The poor agreement of the ABINIT calculations with the experimental results corresponds to the poor agreement found in Fig. \ref{Fig6} for the TA-phonons along the $\Gamma$-X direction.  The reason for the poor agreement found in this case is not known to us in detail. We presume it is due to inaccuracies in the \textit{ab initio} pseudopotentials employed.

\begin{figure}[tbph]
\includegraphics[width=8cm ]{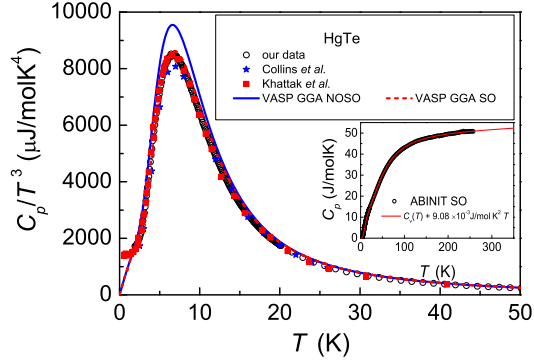}
\caption{$C_v/T^3$ vs. $T$ for HgTe calculated with both, ABINIT-LDA and VASP-GGA codes including and not including \textit{s-o} interaction compared with our experimental data. The inset depicts $C_v$$\approx$ $C_p$ obtained by ABINIT with \textit{s-o} interaction included. A contribution linear in temperature, 9.08$\times$10$^{-3}$ J/mol\,K$^2$ $T$ (Ref. \onlinecite{Landolt}), due to thermal expansion has been added to the calculated data.} \label{Fig14}
\end{figure}

{\color{black} At high temperatures contributions linear in temperature due to thermal expansion have been added to the ABINIT data (see insets in Fig. \ref{Fig13} and Fig. \ref{Fig14}). While the linear term for HgTe (9.08$\times$10$^{-3}$ J/mol\,K$^2$ $T$) is in good agreement with literature data, that for HgSe ($\sim$4$\times$10$^{-3}$ J/mol\,K$^2$ $T$) is smaller by a factor of 3.5 than data reported in the literature.~\cite{Landolt} Our experimental results are, however, in very good agreement with data reported by Gul'tyaev and coworkers.~\cite{Gultyaev1959}}

\begin{figure}[tbph]
\includegraphics[width=8cm ]{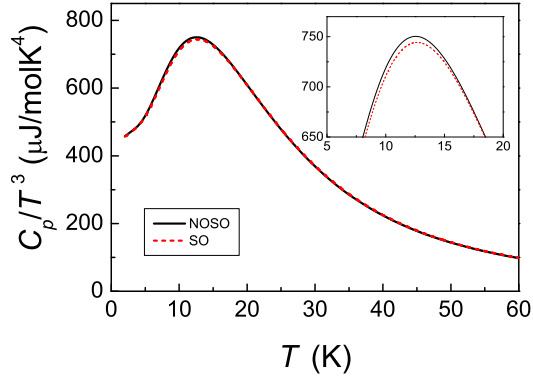}
\caption{(color online) Calculations, using the VASP-GGA code with (red) dashed line and without (solid line) \textit{s-o} interaction, of  HgS at a pressure of $\sim$25 GPa (rock salt structure). Notice that the \textit{s-o} interaction lowers $C_v /T^3$ by an amount that is insignificant with respect to uncertainties in the computational method. The inset has been added in order to clearly illustrate the small effect of \textit{s-o} interaction.
} \label{Fig15}
\end{figure}

We would like to recall the fact that the \textit{s-o} effects calculated here for materials with zincblende structure are opposite to those calculated in Romero \textit{et al.}~\cite{Romero2008} for the lead chalcogenides, which crystallize in the rock salt structure. In order to investigate whether this difference is characteristic of the crystal structures, we calculated  the dispersion relations of  rock salt type HgS which, as shown in Fig. 9, is stable above 21.6 GPa (the experimental phase transition pressure is 20.5 GPa~\cite{Nelmes1998}).   There are, of course, no measurements for HgS in the rock salt phase but we show in Fig. \ref{Fig15} the $C_v/T^3$ vs. $T$ calculated with the VASP-GGA code. In order to stabilize the rock salt structure, we performed the calculations assuming a pressure of 25 GPa applied to the material. The \textit{s-o} interaction induces a minute decrease of $C_v/T^3$ (less than 1\%) which can be assumed to be within the possible error of the calculations using different potentials. The change in the sign of the effect  when going  from zincblende to rock salt, found in the calculations going from the Hg chalcogenides  to the Pb chalcogenides, is not confirmed by the calculations performed for HgS for the rock salt structure.

The difference between zincblende and rocksalt, however,
has the right sign. We thus conclude that the reason for the change in sign of the effect of \textit{s-o} interaction, from the Hg to the Pb chalcogenides is likely to be related to the change in structure.

The reader may wonder why the maximum of $C_v/T^3$ in Fig. \ref{Fig15} is smaller by a factor of seven than calculated for the zb phase. In order to ascertain the reason for this difference we calculated the phonon DOS for the rs phase. We found the TA peak to occur at 62 cm$^{-1}$, a factor of 2.3 larger than that shown in Fig. \ref{Fig12}(b) for the zb phase. This difference can account for the lower maximum in $C_v/T^3$ calculated for the rs phase.

\section{Conclusions}\label{SecSummary}
The original motivation of the present work was to perform accurate measurements of the specific heat of  $\beta$-HgS, HgSe and  HgTe at low temperatures and to compare them with \textit{ab initio} calculations of the band structures of these materials, thought to be semimetals with the $\beta$-tin inverted valence band structure. A prerequisite was the calculation of the electronic band structures of these materials. While doing so, we found a number of thus far unknown details and their interpretation. We mention first that there is a significant hybridization of the \textit{d} core electrons of  Hg with the (chalcogen) \textit{p}-like valence bands which results in a negative contribution to their \textit{s-o} splitting at $\Gamma$. Consequently, for $\beta$-HgS this \textit{s-o} splitting even reverses its sign. Thus, the material, believed to be a semimetal, becomes a semiconductor with a gap similar to the magnitude of the $\Gamma$ \textit{s-o} splitting ($\sim$ 0.1 eV). The \textit{s-o} splittings calculated for the three materials compare satisfactorily with experimental results, whereas the negative $E_0$ gaps are larger in magnitude as a consequence of the so-called gap problem associated with LDA. The presence of \textit{s-o} interaction in our calculations allows us to determine the coefficients of the linear terms of the $\Gamma_8$ valence bands. They compare well with estimates based on a semiempirical interpolation formula used in the past for other zincblende-type semiconductors.

We next proceeded to the calculation of the phonon dispersion relations and comparison with experimental results based on Raman and INS spectroscopies. We performed these calculations using existing computer codes and investigated the effect of \textit{s-o} interaction which turned out, in the case of the acoustic phonons, to have a sign opposite to that found for the rock-salt structure of the Pb chalcogenides. This effect resulted in changes by about 10\% in the values of the calculated maximum of $C_v/T^3$ vs. $T$: the \textit{s-o} splitting lowers the value at this maximum, an effect which is opposite to that found for the rock-salt structure lead chalcogenides. An argument was found to correlate the sign of this effect with the structure of the materials by calculating $C_v/T^3$ for HgS in the rock salt structure. From the calculated phonon dispersion relations we obtained the phonon densities of states and their atomic projections. They were used for an analysis of measured phonon line widths and the corresponding effects of isotopic composition. We concluded that there must be a large contribution of the electron-phonon interaction across the $\Gamma_8$ bands to the line widths of the Raman phonons of HgSe and HgTe. We also calculated two-phonon densities of states in order to interpret extant and future second order Raman spectra. Finally we investigated a few effects of hydrostatic pressure on the Raman phonons of these materials, in particular HgTe. We calculated the Gr\"uneisen parameters of LO and TO phonons and also of the transverse (Born) effective charge. The latter was found to be $\gamma^*$ = -0.65 for HgTe, a value similar to that found for most other zincblende type semiconductors. the sign of $\gamma^*$ has been interpreted as signaling a decrease of the ionicity with increasing pressure.

\begin{acknowledgments}
A.H.R. has been supported by CONACyT Mexico
under project J-59853-F and by PROALMEX/DAAD. Further computer resources
have been provided by CNS IPICYT Mexico.
A. M. acknowledges the financial support from the Spanish MCYT under grants MAT2007-65990-C03-03,  CSD2007-00045 and the supercomputer  resources provides by the Red Espa\~{n}ola de Supercomputaci\'{o}n.
We thank E. Br\"ucher for the magnetic measurements and V. Duppel for the microprobe analysis of the natural $\beta$-HgS crystals. We are also indebted to K. Syassen for a critical reading of the manuscript and to N. E. Christensen for pointing out to us references ~\onlinecite{Rohlfing1998,Fleszar2005,Carrier2004}.

\end{acknowledgments}

\end{document}